\documentclass[12pt]{article}
\usepackage{graphicx}

\usepackage{amsmath}

\usepackage{verbatim}

\usepackage[hidelinks]{hyperref}

\usepackage{cite}

\usepackage[font=small,skip=6pt]{caption}

\newcommand\pubnumber{SNSN-323-63}
\newcommand\pubdate{\today}

\def\institute{University of Belgrade\\
Institute of Physics Belgrade, Pregrevica 118, Zemun, SERBIA}
\def\support{\footnote{Work supported by the Swiss National Science Foundation, Switzerland and
Ministry of Education, Science and Technological Development, Serbia under the Project 171019.}}

\def\Title#1{\begin{center} {\Large #1 } \end{center}}
\def\Author#1{\begin{center}{ \sc #1} \end{center}}
\def\Address#1{\begin{center}{ \it #1} \end{center}}

\newcommand\pubblock{\rightline{\begin{tabular}{l} \pubnumber\\
         \pubdate  \end{tabular}}}
\newenvironment{Abstract}{\begin{quotation}  }{\end{quotation}}
\newenvironment{Presented}{\begin{quotation} \begin{center} 
             PRESENTED AT\end{center}\bigskip 
      \begin{center}\begin{large}}{\end{large}\end{center} \end{quotation}}
\def\Acknowledgements{\bigskip  \bigskip \begin{center} \begin{large}
             \bf ACKNOWLEDGEMENTS \end{large}\end{center}}

\def\beq{\begin{equation}}
\def\eeq#1{\label{#1}\end{equation}}
\def\eeqn{\end{equation}}

\def\beqa{\begin{eqnarray}}
\def\eeqa#1{\label{#1}\end{eqnarray}}
\def\eeqan{\end{eqnarray}}

\let\bar=\overbar

\def\Dslash{\not{\hbox{\kern-4pt $D$}}}
\def\dslash{\not{\hbox{\kern-2pt $\del$}}}

\def\msb{{\bar{\ssstyle M \kern -1pt S}}}

\newcommand{\pt}{p_{T}}

\newcommand{\Hbb}{{H \rightarrow b \bar{b}}}

\newcommand{\TeV}{\,\mathrm{TeV}}

\newcommand{\pb}{\,\mathrm{pb}}

\newcommand{\ifb}{\,\mathrm{fb^{-1}}}

\newcommand{\ttbar}{{t\bar{t}}}

\newcommand{\Hut}{t \rightarrow Hu}
\newcommand{\Hct}{t \rightarrow Hc}

\begin{document}
\begin{titlepage}
\pubblock

\vfill
\Title{Search for flavor-changing interactions of the top quark with the Higgs boson in $\Hbb$ channel at $\sqrt{s}\,=\,13\TeV$}
\vfill
\Author{ Predrag Cirkovic\support~on behalf of the CMS Collaboration}
\Address{\institute}
\vfill
\begin{Abstract}
A search for flavor-changing neutral current (FCNC) processes in associated production of a top quark and Higgs boson, with the Higgs boson decaying to a pair of b quarks, is presented.
To experimentally probe the top-Higgs FCNC couplings, for the first time the single-top production is considered as a signal process.
One isolated lepton and at least three reconstructed jets, among which at least two are identified as b quark jets, are found in the final state.
The data sample corresponds to an integrated luminosity of $35.9\ifb$ recorded by the CMS experiment at the LHC in proton-proton collisions at $\sqrt{s}\,=\,13\TeV$ in 2016, and the final results are presented in the form of the observed and expected 95\% CL upper limits on the branching ratio of top quark decays.
\end{Abstract}
\vfill
\begin{Presented}
$\mathrm{10^{th}}$ International Workshop on Top Quark Physics\\
Braga, Portugal, September 17--22, 2017
\end{Presented}
\vfill
\end{titlepage}
\def\thefootnote{\fnsymbol{footnote}}
\setcounter{footnote}{0}

\section{Motivation}

In the Standard Model (SM), flavor-changing neutral currents (FCNC) are forbidden at tree level and strongly suppressed in loop corrections~\cite{Glashow:1970gm}.
Several extensions of the SM incorporate significantly enhanced FCNC that can be directly probed at the LHC~\cite{AguilarSaavedra:2004wm}.
This report summarizes the latest CMS~\cite{Chatrchyan:2008aa} results of the top-Higgs FCNC analysis with $H \rightarrow b\bar{b}$ decays performed for the associated production of a top quark with a Higgs boson i.e. $gu/c \rightarrow t(\rightarrow \ell^+ \nu b)H(\rightarrow b\overline{b})$ (abbr. ST, used for the first time in this analysis) and the FCNC decays of top quarks in $\ttbar$ semileptonic events i.e. $gg \rightarrow t(\rightarrow \ell^+ \nu b) \overline{t}(\rightarrow {\rm \overline{u}/\overline{c}}H(\rightarrow b\overline{b}) )$ (abbr. TT)~\cite{CMS:2017cck}.
The SM branching fraction of $t \rightarrow H q$ is predicted to be $\mathcal{O}(10^{-15})$~\cite{AguilarSaavedra:2008zc}.
The predicted cross section at $13\TeV$ for ST under the assumption $\kappa_{Hut}=1\,,\kappa_{Hct}=0$~($\kappa_{Hut}=0\,,\kappa_{Hct}=1$) is $13.84~(1.90)\pb$, while for the TT, it is $36.98\pb$ for each of the couplings.
The cross section times branching ratio is calculated for both signatures.
The sensitivity to $\kappa_{Hut}$ coupling can be improved by exploiting both the TT and ST processes.

\section{Analysis Strategy}

The events with exactly one isolated lepton (electron or muon) and at least two identified b jets are considered.
The Monte Carlo (MC) simulation of the signal and the dominant background (the SM $\ttbar$) is done at leading order (LO) using MADGRAPH~\cite{Alwall:2014hca}, and the next-to-leading order (NLO) using POWHEG v2~\cite{Alioli:2011as}, respectively.
The particle-flow (PF) algorithm~\cite{Sirunyan:2017ulk} is used to reconstruct and identify individual particles.
Jets are reconstructed by clustering PF candidates using the anti-kt~\cite{Cacciari:2008gp} algorithm with a distance parameter of 0.4.
Tracks are assigned to vertices according to CMS tracking and vertex finding algorithms and combined with photon candidates.
The jet finding algorithm is applied to these physics objects.
Single lepton events are recorded using a trigger passing at least one electron (muon) with $p_T >$~32~GeV (24~GeV) selected within $|\eta| <$~2.1.
Lepton candidates are selected requiring $|\eta| <$~2.1 with $p_T >$~35~GeV (30~GeV).
Leptons are isolated using a relative isolation requirement $I_{rel} <$~0.06 (0.15) and corrected for energy deposits from pileup interactions.
Multilepton background processes are suppressed rejecting events with additional leptons passing the looser isolation requirement of $I_{rel} <$~0.25 and $p_T > $~10~GeV.
The events must have at least three jets with $p_T >$~30~GeV within the $|\eta| <$~2.4.
As signal events contain three b quarks in the final state, the requirement for the presence of at least three jets in event with at least two of them identified as b jets is applied.
The identification (b-tagging) is performed using ''Combined Secondary Vertex v2'' (CSVv2) algorithm using the medium working point~\cite{CMS-PAS-BTV-15-001}.
The signal sensitivity is increased by splitting events into five categories based on the total number of reconstructed jets and the number of b tagged jets.
A full kinematic reconstruction of the event is performed for several hypotheses: ST, TT, and $\ttbar$ semileptonic background event.
It is done for all possible permutations of the b jets to be associated with the Higgs boson and the top quark decay products.
The reconstructed kinematic variables are used as the input variables of the multivariate analysis (MVA) that uses a Boosted Decision Tree (BDT) algorithm.
The b jet permutation with the highest BDT score is chosen as the correct one.
\begin{figure}[htb!]
\centering
\includegraphics[width=7.0cm,clip,angle=0,origin=c]{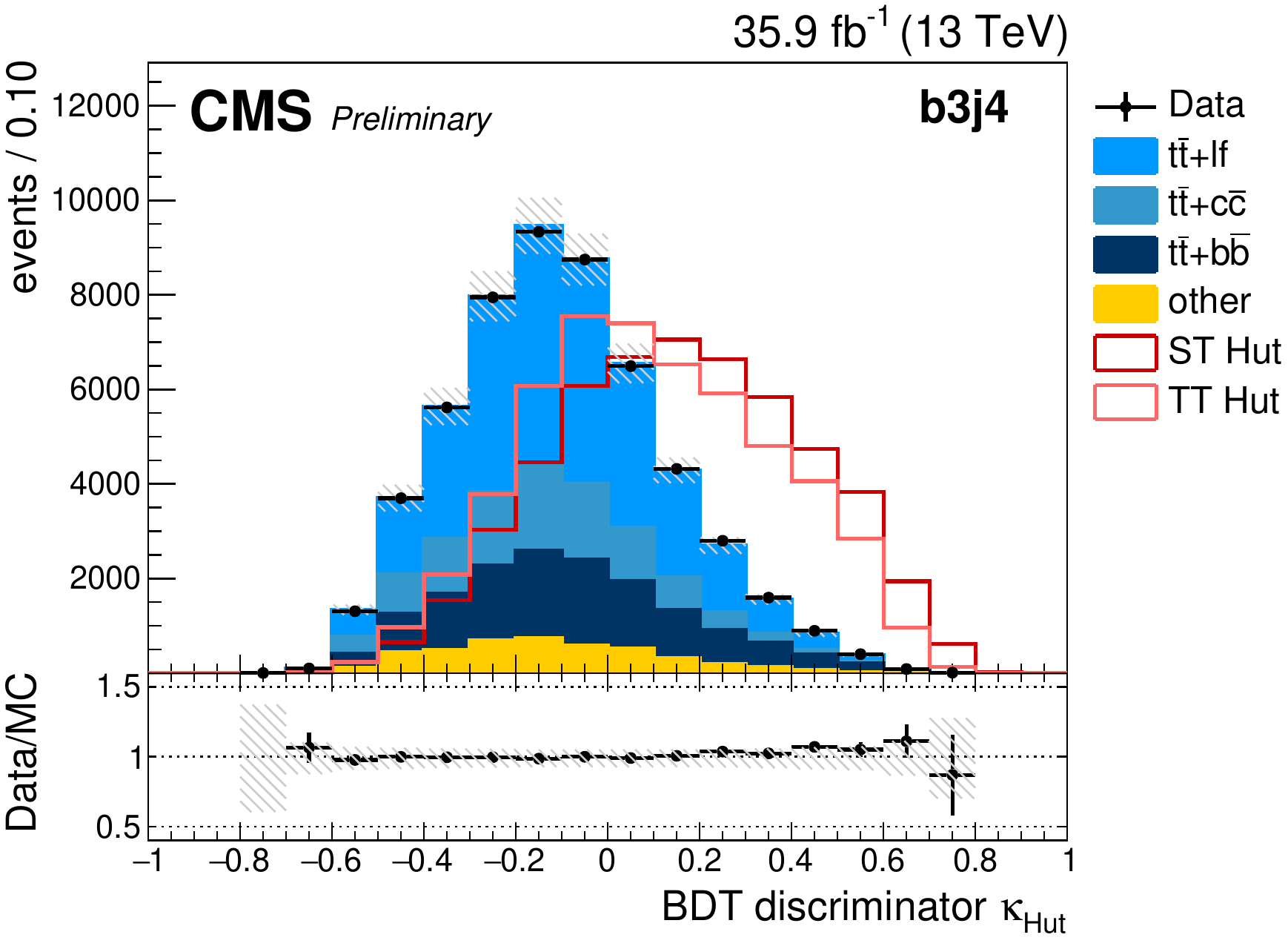}
\qquad
\includegraphics[width=7.0cm,clip,angle=0,origin=c]{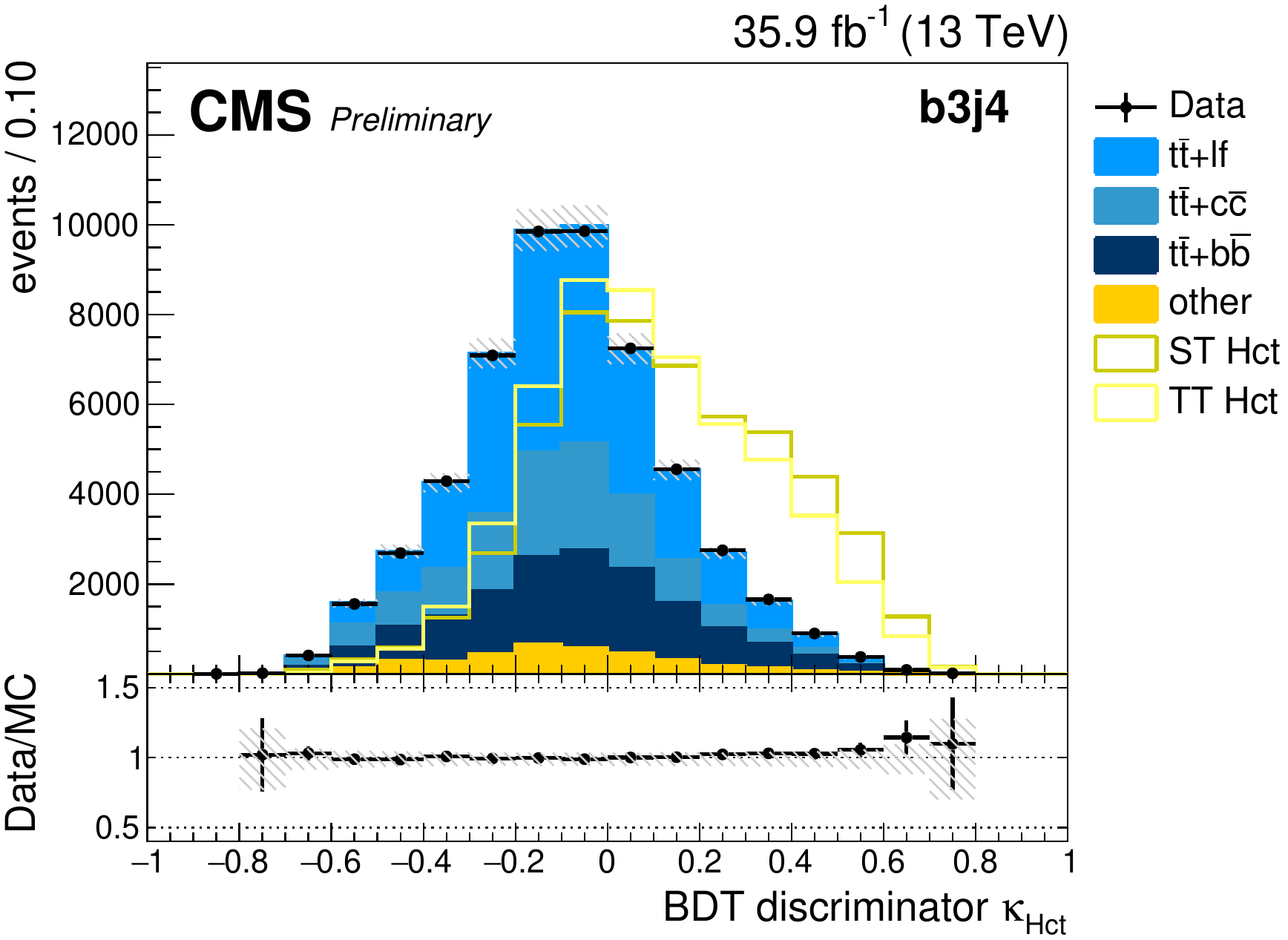}
\caption{
BDT output score distributions for b3j4 category for $\Hut$ (left) and $\Hct$ (right) training after the fit to data.
All background processes are constrained to the SM expectation in the fit.
The signal distributions are normalised to the total number of events in the predicted backgrounds.
The shaded area corresponds to a total uncertainty on the predicted background~\cite{CMS:2017cck}.
}
\label{fig:MVAHutHct_b3j4}       
\end{figure}
For each jet category, the generated $t \rightarrow u/cH$ signal events are trained against the sum of all background events in a BDT.
The variables with the best separation power are: the lepton charge (considered only $t \rightarrow uH$), the CSVv2 of the lowest $\pt$ b jet from Higgs boson decay, the reconstructed mass of the Higgs boson, and the BDT score used for the b jet assignment.
In each $t \rightarrow u/cH$ jet category, the corresponding BDT score distribution is used for the signal extraction.
Fig~\ref{fig:MVAHutHct_b3j4} shows the data-MC comparison for this observable after the fit to data.
Various sources of systematic uncertainty that affect both the normalisation and shape of the predicted background are taken into account as nuisance parameters in the computation of the exclusion limits~\cite{CMS:2017cck}.
The systematic uncertainties considered in the simulation are independent variation of the factorisation and renormalisation scales, $\mu_{F}$ and $\mu_{R}$, and variation in the PDF and $\alpha_{S}$.
The dominant systematic uncertainty arises from the application of the b tagging requirement.

\section{Results}

\begin{figure}[hbt!]
  \begin{center}
    \includegraphics[width=7.0cm]{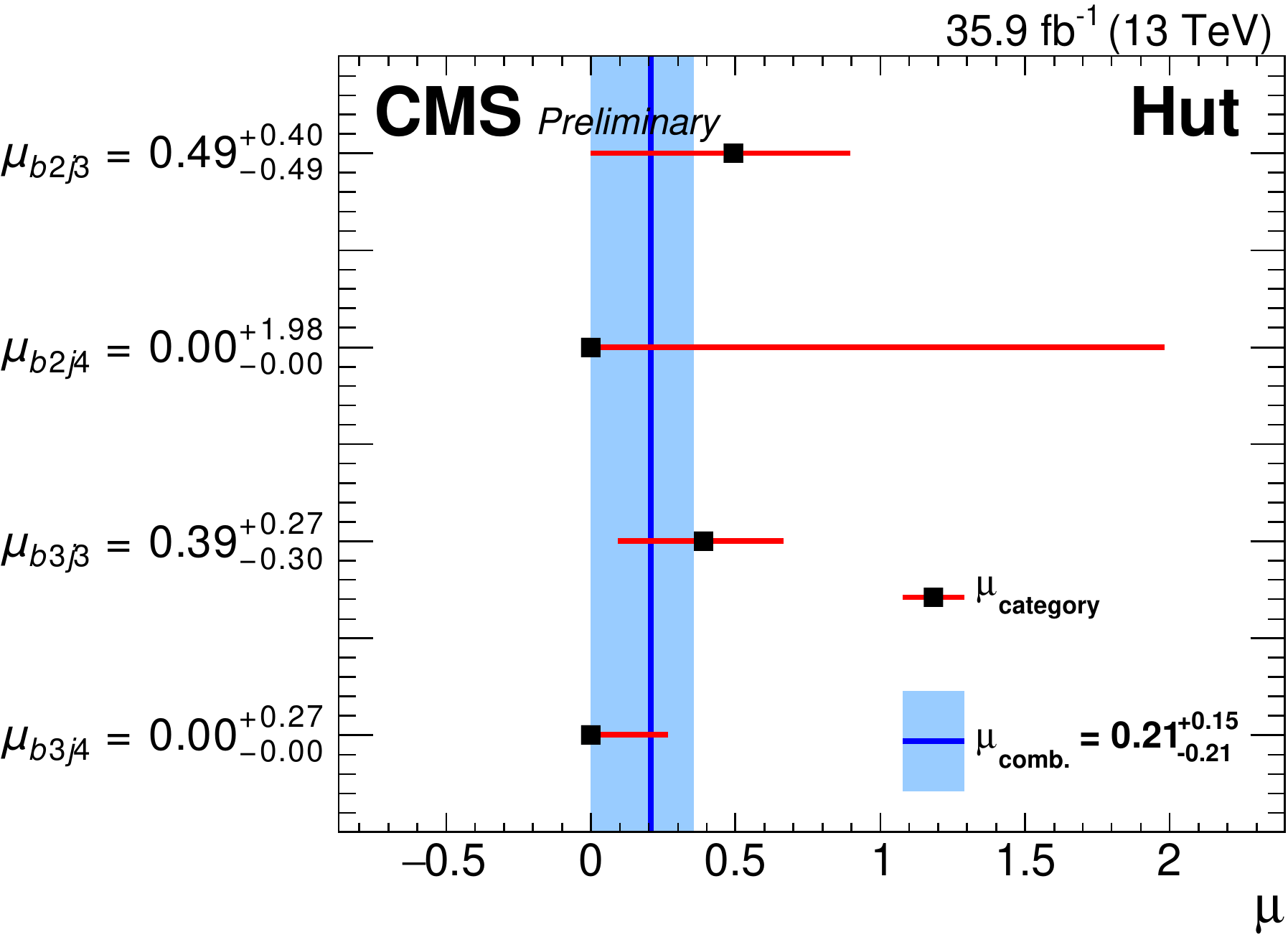}
	  \quad
    \includegraphics[width=7.0cm]{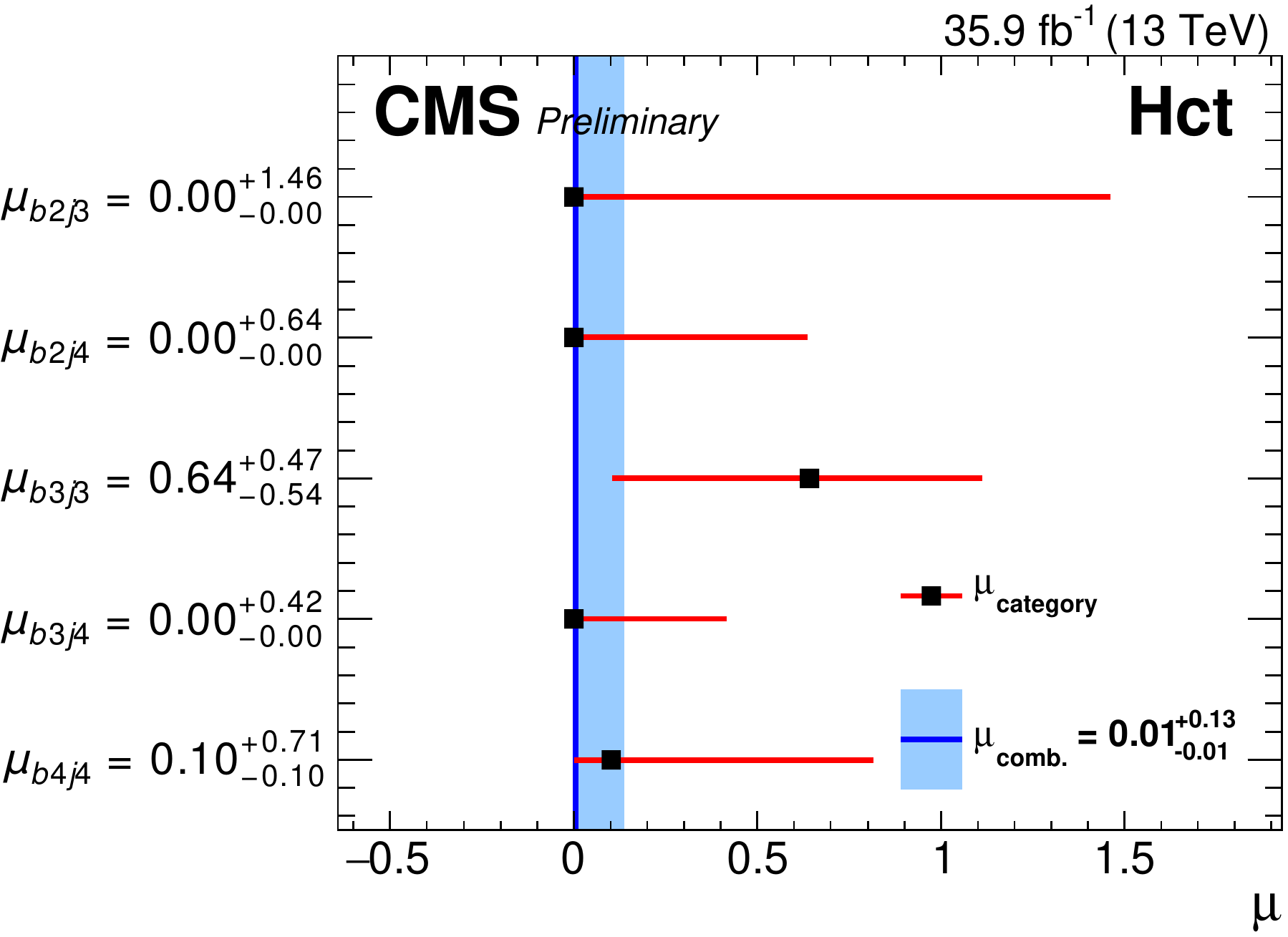}
    \caption{The best fit signal strength for Hut (left) and Hct (right).
		The signal strength is restricted to positive values in the fit~\cite{CMS:2017cck}.}
    \label{fig:BestFit}
		\vspace{3pt}
    \includegraphics[width=7.0cm, angle=0]{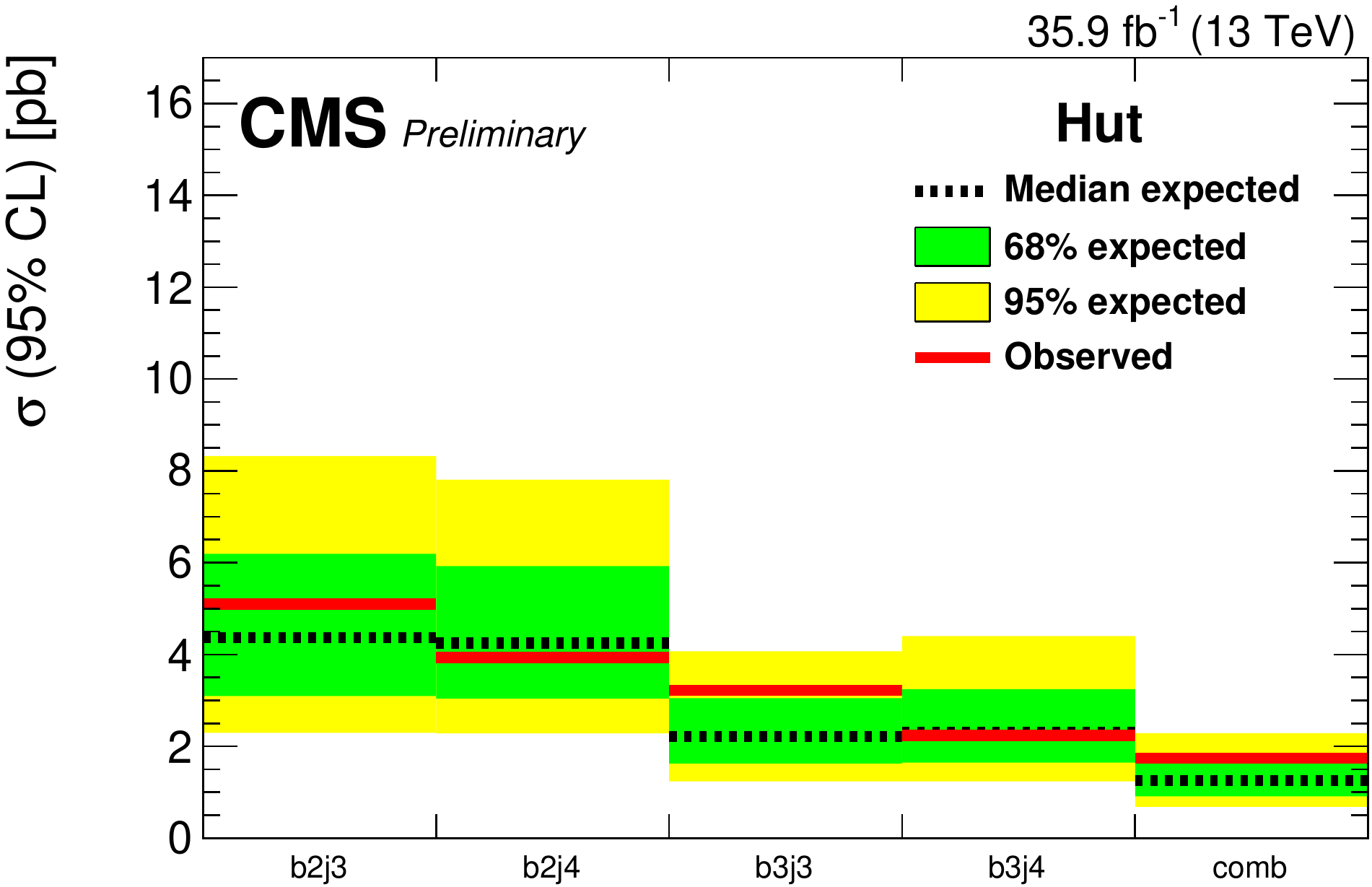}
		\quad
    \includegraphics[width=7.0cm, angle=0]{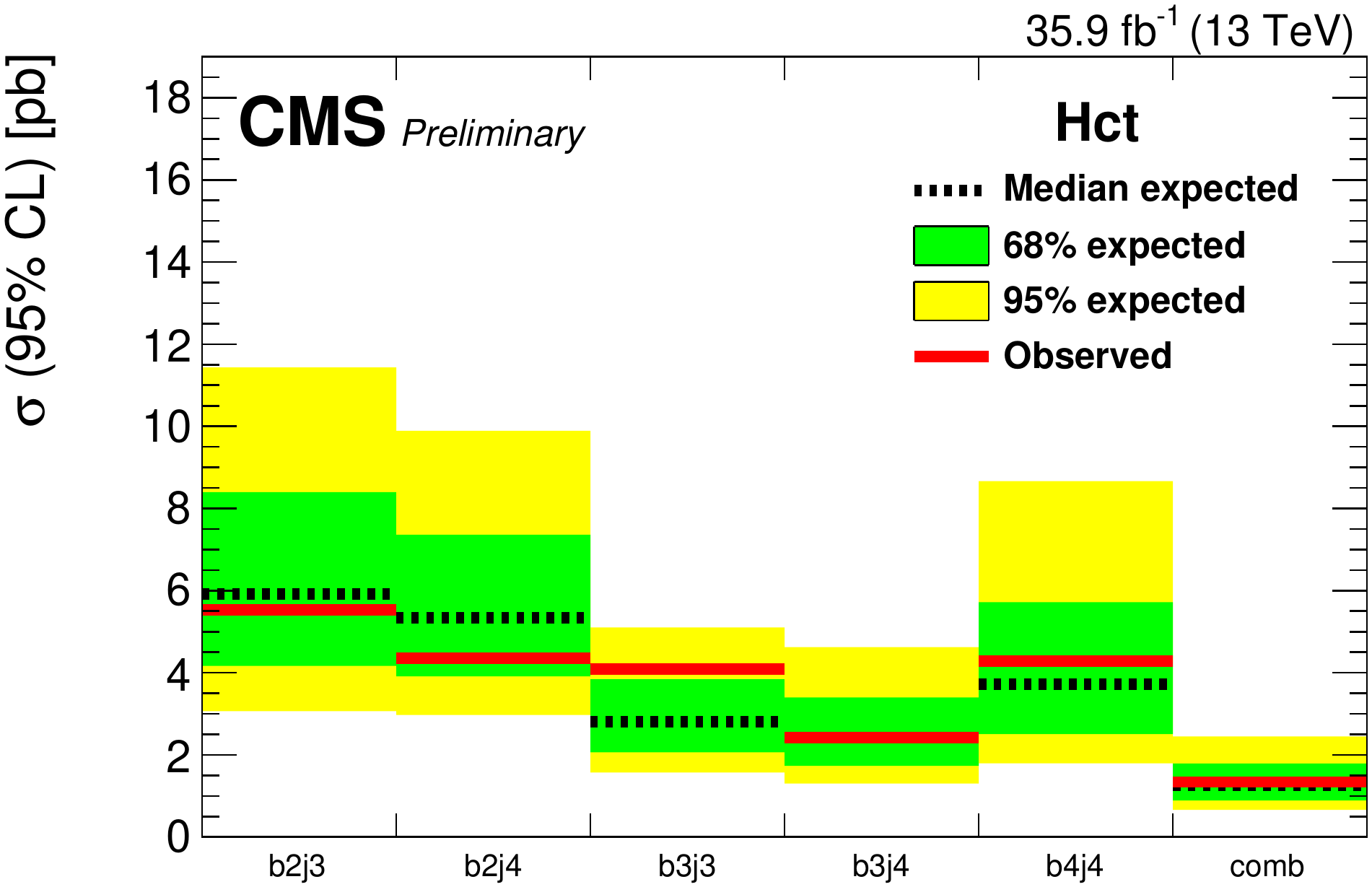}
		\\\vspace{3pt}
		\includegraphics[width=7.0cm, angle=0]{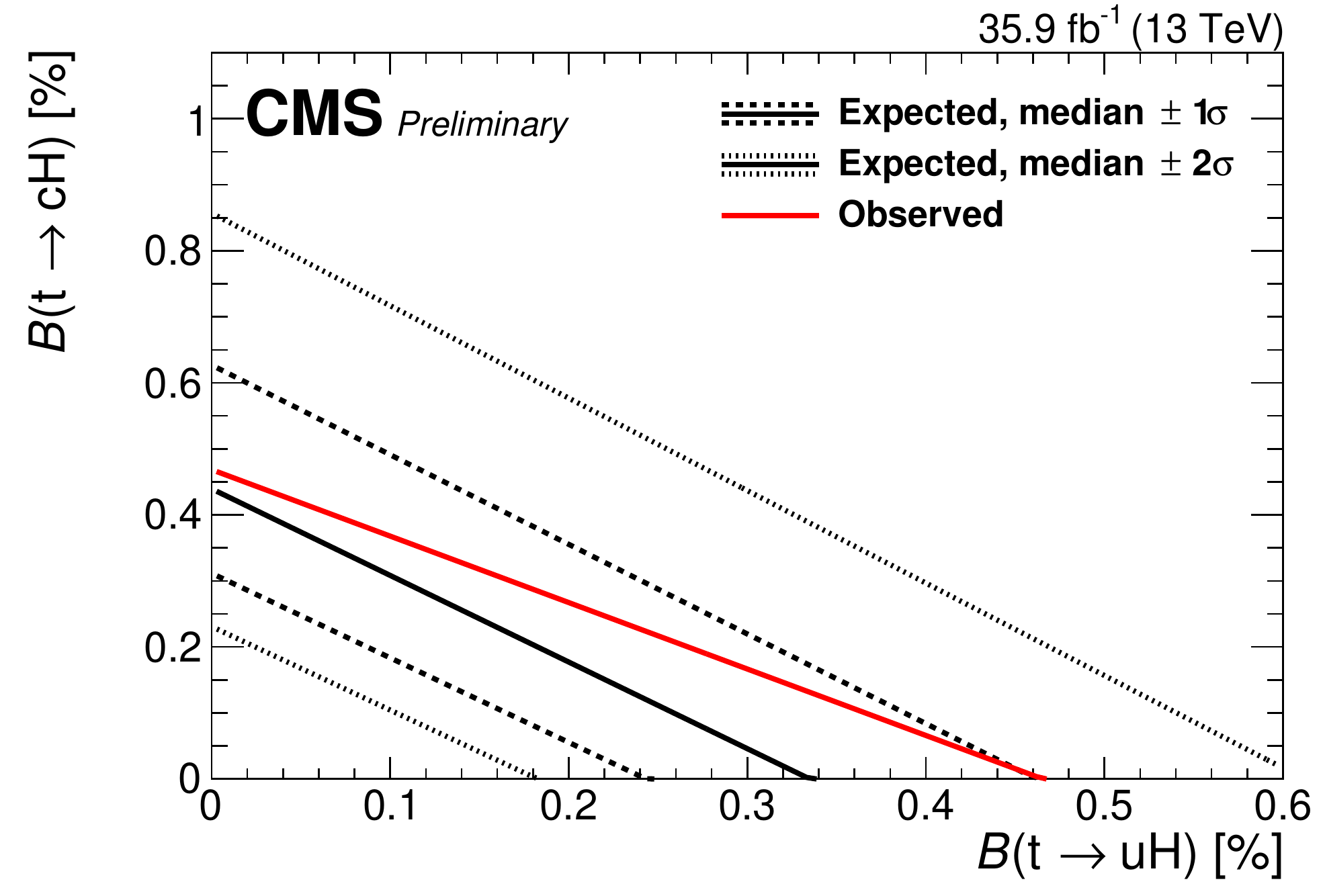}
    \caption{Excluded signal cross section at 95\% CL per event category for $t \rightarrow uH$ (upper-left, where the b4j4 category is excluded due to negligible improvement in the final sensitivity) and Hct (upper-right).
		Upper limits on $B(t \rightarrow u H)$ and $B(t \rightarrow c H)$ at 95\% CL (bottom)~\cite{CMS:2017cck}.}
    \label{fig:Limits}
  \end{center}
\end{figure}

The final results are presented as the exclusion limits set on the top quark FCNC decay branching ratios.
A 95\% confidence level (CL) upper limit is computed for the production cross section of top-Higgs FCNC events times branching ratios of the top quark semileptonic decay and the Higgs boson decay to b quarks~\cite{CMS:2017cck}.
The expected and observed limits on the signal production cross section are computed per each category, as well their combination (Fig~\ref{fig:Limits}).
A simultaneous fit is performed to all categories and the resulting signal strength is shown in Fig~\ref{fig:BestFit}.
The observed (expected) 95\% CL exclusion limits on the branching ratios $B(t \rightarrow u H) <$~0.47(0.34)\% and $B(t \rightarrow c H) <$~0.47(0.44)\% of the top quark FCNC decays are obtained.
The results are also interpreted for the scenario where both FCNC couplings have non-zero values (Fig~\ref{fig:Limits}, bottom).

\section{Conclusion}

The results of the search for the top-Higgs FCNC with the $35.9~\mathrm{fb}^{-1}$ data collected in proton-proton collisions at $\sqrt{s}=13\TeV$ are presented.
This search considered $\ttbar$ pair production with a top quark FCNC decay, together with production of single-top FCNC in association with a Higgs boson, which is used for the first time.
The observed (expected) $95\%$ CL upper limits have been set on the branching ratios of top quark decays, $B(t \rightarrow u/c H) <$~0.47/0.47\%~(0.34/0.44\%).
Including the single-top FCNC production mode in this study, the upper limit on $t \rightarrow uH$ branching ratio has been improved by $\simeq$~20\% relatively to the exclusive usage of the $\ttbar$ FCNC process.

\Acknowledgements
Thanks to the CMS Collaboration for running the experiment and for all the help during the analysis and preparation of presentation,
the CERN accelerator division for the excellent operation of the LHC,
and SCOPES and MPNTR for supporting this work.

\end{document}